\documentclass[aps, prb, letterpaper, 10pt, twocolumn, superscriptaddress]{revtex4-2}
\usepackage[colorlinks=true,allcolors=blue!70!black]{hyperref}
\usepackage{amssymb}
\usepackage{amsmath}
\usepackage{graphicx}
\usepackage{subfigure}
\usepackage{amsfonts}
\usepackage{xcolor}
\usepackage{epsfig}
\usepackage{color}
\usepackage{bbm}
\usepackage{bm}
\usepackage{tabularx}
\usepackage{multirow}
\usepackage{mathtools}
\usepackage{xfrac}
\usepackage{physics}
\usepackage{enumerate}
\usepackage{braket}
\usepackage{booktabs}
\usepackage{siunitx,isotope}

\newcommand{\mean}[1]{\langle #1 \rangle}
\newcommand{\one}{\openone}

\newcommand{\hatd}[1]{\hat{#1}^{\dagger}}

\newcommand{\half}{\frac{1}{2}}

\newcommand{\vnnb}{V$_{\rm B}$}

\begin{document}
\title{Decoherence time of the ground state spin of \vnnb\ centers in hexagonal boron nitride}

\author{F. T. Tabesh}
\email{fatemeh.tabesh@gmail.com}
\affiliation{School of Quantum Physics and Matter, Institute for Research in Fundamental Sciences (IPM), P.O. 19395-5531, Tehran, Iran}
\affiliation{School of Physics, Institute for Research in Fundamental Sciences (IPM), P.O. 19395-5531, Tehran, Iran}
\author{S. Rahimi-Keshari}
\email{s.rahimik@gmail.com}
\affiliation{School of Quantum Physics and Matter, Institute for Research in Fundamental Sciences (IPM), P.O. 19395-5531, Tehran, Iran}
\affiliation{School of Physics, Institute for Research in Fundamental Sciences (IPM), P.O. 19395-5531, Tehran, Iran}
\author{M. Abdi}
\email{mehabdi@gmail.com}
\affiliation{Wilczek Quantum Center, School of Physics and Astronomy, Shanghai Jiao Tong University, Shanghai 200240, China}
\begin{abstract}
The ground-state spin of optically active defects in hexagonal boron nitride (hBN) offers a promising platform for quantum information applications, such as qubits for quantum computing and nanoscale sensing. 
A key characteristic of a qubit is its decoherence time, as its duration and controllability are critical for practical applications in quantum technologies. In this work, we investigate the electron spin dephasing time of the negatively charged boron vacancies, \vnnb\ centers, in the hBN lattice by considering the dipolar hyperfine as well as spin-phonon interactions. We employ an approximate method based on the Holstein-Primakoff transformation to take into account a large number of nuclear spins and Debye model to consider the effect of lattice phonons. We show that, in the presence of the dipolar hyperfine interactions, Hahn-echo coherence time of the \vnnb\ electron spin is approximately $30\: \mathrm{\mu s}$ at room temperature. 
Our results provide a step forward in understanding  the \vnnb\ defect decoherence in the hBN, which might be used for quantum information applications.

\end{abstract}

\maketitle

%
%
\section{Introduction}%
One of the promising platforms in the field of quantum technology is color centers (optically active defects in solids), which can be used as solid qubits for quantum computing and quantum sensing \cite{awschalom2018, Bradley2019, Weber, atature2018, waldherr2014}.  
Color centers in bulk semiconductors, such as nitrogen-vacancy centers in diamond \cite{Doherty2013, Schirhagl} and various defects in silicon carbide \cite{Riedel, koehl2011room, Castelletto_2020} are of particular interest for quantum technological applications. These color centers can operate at room temperature and exhibit millisecond coherence time~\cite{balasubramanian2009, widmann2015}. However, accessing color centers in bulk samples can be practically challenging. Moreover, for quantum nanoscale sensing, it is desirable for the sensors to be in close proximity to the target to maximize sensitivity. However, defects located near the surface of bulk materials experience degraded coherence times~\cite{Zhang2017}. To overcome these limitations, researchers explore alternative platforms, such as solid-state point defects in two-dimensional (2D) materials or van der Waals (vdW) crystals \cite{liu20192d, liu2016van, Novoselov}, which offer promising advantages for nanoscale sensing and quantum technologies.

Recent studies show that the most promising candidate of the 2D material is hexagonal boron nitride (hBN) owing to its wide bandgap (${\sim}6\  \mathrm{eV}$)~\cite{Cassabois2016, Tran2016}. The electronic and spin properties of various defects in hBN have been extensively studied theoretically. In particular, negatively charged boron vacancy defects (\vnnb)~\cite{Abdi2018,Ivady2020} has attracted significant attention due to its well-known molecular structure, ease of generation, and suitability as a single-photon emitter at room temperature~\cite{Tran2016,Sajid2020}. Additionally,  it can be integrated into advanced cavities and resonators and used as a versatile, high-resolution 2D quantum sensor~\cite{healey2021q, Tetienne2021b,Tabesh2023, Tabesh_2022}. Recent experiments have also demonstrated that the \vnnb\ defect can be easily prepared, controlled, and optically read out~\cite{Gottscholl2020, Kianinia2020}.

A key characteristic of these defects is their decoherence time $T_2$, which is crucial for quantum technological applications. For instance, in quantum sensing applications, the sensitivity is proportional to $1/\sqrt{T_2}$ \cite {Degen2017}. In addition, to achieve quantum error correction $T_2$ must be on the order of 100 $\mu$s, given a time scale of ${\sim}$ 10  ns for gate operation~\cite{Graham, divincenzo2000}. Therefore, various studies have recently investigated the coherence time of defects in hBN theoretically and experimentally; see Ref.~\cite{Liu_2022} for a review. For example, Ref~\cite{ye2019spin} has reported the coherence time of 36 $\mu$s for a qubit at the vacancy site of nitrogen–substitution–vacancy complex $N_B V_N$ using a cluster expansion method, and has also shown that $T_2$ increases monotonically as all \isotope[11]{B} nuclei are replaced by \isotope[10]{B} nuclei.

Various coherence times have been reported for the \vnnb\ defect in hBN. By applying pulsed spin resonance protocols, the \vnnb~spin coherence time of  $2\: \mu s$ has been measured at room temperature, and the upper limit of the inhomogeneous spin dephasing time $T_2^* = 100 ns$ estimated at $8.5 m\mathsf{T}$ magnetic field, induced by surrounding magnetic moments~\cite{Gottscholl2021}. Subsequently, the spin coherence time of  \vnnb~ was estimated to be of the order of $10\: \mu s$  at $ \mathrm{T} = 8\: \mathrm{K}$, with a reported inhomogeneous spin dephasing time of $T_2^*=2.7\: \mu s$~\cite{gottscholl_2020n}. Numerical simulations employing cluster correlation expansion methods and considering the Fermi contact and the dipolar interactions reported $T_2 = 92\: n s$ for the h$^{11}$BN crystal and $T_2 = 115 \; ns$ as well as inhomogeneous spin dephasing time $ T_2^*=20\: ns $ for the h$^{10}$BN crystal by applying a static magnetic field $B \sim 15\:m\mathsf{T}$~\cite{haykal2022}. Further studies demonstrated coherent coupling of the \vnnb\ defect with three neighboring nitrogen atoms due to nuclear quadrupole interaction, reporting a Hahn-echo coherence time of  $ T_{\rm coh}=15 \mu s$. In addition, several studies have focused on increasing the $T_2$ coherence time of the \vnnb\ defect using various methods. It was shown that applying a strong continuous microwave drive with modulation to stabilize Rabi oscillations extends the coherence time to approximately $\sim 4 \mu s$ ~\cite{ramsay2022room}. Also, a combination of isotopic enrichment and strain engineering has been shown to significantly enhance $T_2$, yielding $207.2 \mu s$ and $161.9 \mu s$ for single- and multi-layer h$^{10}$BN lattice, respectively~\cite{lee2022first}. Additional values for inhomogeneous  spin dephasing times and coherence times have also been reported, as summarized in Table \ref{table}.

\begin{table}[t]  
    \centering 
\begin{tabular}{SSSSS} \toprule
    {Reference} & {$T_2$} & {$T^{*}_{2}$} & {$B$} & {$\textsl{T}$}  \\ \midrule
     {Ref.~\cite{Gottscholl2021}} & {$2\: \mu s $}  &{$ 100\: n s $} & {$ 8.5\:m\mathsf{T}$} & {RT} \\
    {Ref.~\cite{gottscholl_2020n}}  & {$10\: \mu s $}  & {$2.7\: \mu s $} & {$-$} & {8\:K} \\
    {Ref.~\cite{haykal2022}} & {$115\: ns\, (^{10}\text{B}) $} & {$20\:ns$}& {$ 15\: m\mathsf{T} $}  & {RT}\\ 
    {Ref.~\cite{haykal2022}} & {$92\: ns\, (^{11}\text{B})$} & {$-$} & {$ 15\: m\mathsf{T} $}  & {RT}\\
   {Ref~.\cite{Murzakhanov}} & {$ 15\: \mu s $}   &  {$-$} &   {$255\: m\mathsf{T} $} & {50\:K}  \\
    {Ref.~\cite{ramsay2022room}}  & {$100\: n s$}  & {$60\:ns$}&  {$ 20\: m\mathsf{T}$ } & {RT}\\
     {Ref.~\cite{lee2022first}}  &  {$45.9\:\mu s$ } &{$-$} & {$3\:\mathsf{T}$}  & {$-$} \\
     {Ref.~\cite{baber2022}}  &  {$-$}  & {$ 19\: n s$}& {$-$}   &{RT}\\
          {Ref.~\cite{rizzato2023extending}}  & {$60\: n s$}  & {$-$} & {$ 8\: m\mathsf{T} $}  & {RT} \\
          {Ref.~\cite{liu2022coherent}}  & {$82\: n s$}  &  {$-$}&  {$ 0 \:m\mathsf{T} $}  &{RT} \\
   {Ref.~\cite{Gao2021}}  & {$1.1\:\mu s$}  & {$120 \:n s$}&{$ 13 \:m\mathsf{T}$}  & {RT} \\ 
   {Ref.~\cite{gao2022nuclear}}  &  {$-$} & {$100\: n s$ } &  {$ 74\: m\mathsf{T} $} & {RT} \\
   \bottomrule
\end{tabular}
\caption{Summary of reported decoherence times. The parameters are defined as follows: $T_2$ is the coherence time, $T^{*}_{2}$ is the inhomogeneous dephasing time, $B$ is the magnetic field, $T$ is the temperature, and RT denotes room temperature.}
\label{table}
\end{table}

In this paper, we focus on two primary sources of dephasing: (i) hyperfine (hf) interactions with the nuclear spin bath of surrounding nuclei and (ii) thermal vibrations of the host material. We assume that the two sources operate independently, and the effects of nuclear electric quadrupole interactions and Fermi contact hyperfine interactions can be ignored. To investigate the effects of the first source, we employ the Holstein-Primakoff (HP) transformation~\cite{Holstein1940}, a numerical bosonic approximation that facilitates the calculation of the electron spin coherence time in a monolayer hBN lattice. This approach mitigates the computational challenges typically associated with large spin systems. For the second source, we model the interaction between the spin degrees of freedom and lattice phonons using the Debye approximation. Specifically, a quantum master equation is introduced to describe the dynamics of the electronic spin under the influence of the phononic bath. By combining the effects of these two independent dephasing mechanisms, we derive the overall decoherence function.

Note that there are additional sources that influence the qubit decoherence~\cite {Doherty2012,Doherty2013,Candido}.
Exploring and understanding the factors affecting spin qubit dynamics—beyond magnetic noise and
lattice vibrations—remains an active and important area
of research.

This paper is organized as follows. In Sec.\ref{spinbath}, we study the coherence time of the \vnnb~ defect by introducing a nuclear spin bath model and employing a hybrid Gaussian approach. To benchmark the performance of our approximate bosonic Gaussian method based on the Holstein-Primakoff (HP) transformation, we compare the numerical results with exact solutions for systems containing a small number of nuclei. We demonstrate that the bosonic approach is in good agreement with exact numerical simulations when the interaction between the central spin and the spin bath is weak. In Sec.\ref{phbath}, we present the central spin reduced master equation to account for interactions with phonons in the hBN lattice, assuming the validity of the Debye approximation. Finally, we summarize our findings in Sec.\ref{con}.

%
%
\section{The nuclear spin bath}\label{spinbath} %
Hexagonal boron nitride is an excellent van der Waals crystal for hosting optically active defects, with each lattice site occupied by either a nitrogen or boron nucleus, both of which possess a non-zero spin. The \vnnb~ center features a triplet spin ground state ($S=1$) with zero-field splitting $D/2\pi\approx 3.5$~GHz. Here, we examine the decoherence time of the \vnnb~defect as influenced by the nuclear spin bath. Firstly, we outline the mechanism that can contribute to the coupling between electron spin $\bf S$ and nuclear spin $\bf I$.  The main and long-range mechanism that has the dominant effect in the spin decoherence is the dipole-dipole interaction, which in analogy to the classical dipolar interaction between magnetic moments its Hamiltonian is written as
\begin{equation}
\hat H_{\rm dd}^{e}=\frac{\hbar\, \mu_0 \gamma_e}{4\pi}\sum_i\frac{\gamma_{i}}{r_i^3}\Bigg[\hat{\bf S}\cdot\hat{\bf I}_i -\frac{3\big(\hat{\bf S}\cdot\mathbf{r}_i\big)\!\big(\hat{\bf I}_i\cdot\mathbf{r}_i\big)}{r_i^2}\Bigg],
\label{hamdd}
\end{equation}
where $\mu_0$ is the permeability of the vacuum ($4\pi\times 10^{-7}$~N$\cdot$A$^{-2}$) and $\hat{\mathbf{r}}_i$ is the displacement unit vector pointing from electron to the $i$th nuclear (in meters). Here, $\gamma_e$ and $\gamma_{n,i}$ are the gyromagnetic ratios of electron and nucleus at site $i$, respectively. The boron and nitrogen gyromagnetic ratios are $\gamma_{\rm B}/2 \pi = 13.66$~MHz/T and $\gamma_{\rm N}/2\pi = 3.078$~MHz/T, respectively. The dipole-dipole interaction depends on the relative orientation of the magnetic moments and thus is anisotropic. 
Pure dipolar interaction is expected if the electron spin is located in a molecular orbital with no overlap with the nuclei.

We consider the nuclear spin lattice of a mono-layer of hBN comprising  equal amounts of boron and nitrogen atoms. Given the non-Markovian nature of a nuclear spin bath, a rigorous full quantum model is necessary to address nuclear spin flip-flops. We also assume an isotopically purified sample, focusing on the most prevalent isotope, \isotope[11]{B}, which accounts for $80\%$ of natural boron. The nuclear spins of the different constituents of the lattice are as follows
\begin{equation}
 I_{\rm N} = 1 \qquad I_{\rm B} = 3/2.
\end{equation}
All the spins interact among each other via dipole-dipole interactions of the form
\begin{align}
\label{eq:dipHam}
\hat{H}_{\rm dd}^{ij} =  g_{ij} \Big[ \hat{\bf I}_i\cdot\hat{\bf I}_j - 3 \big(\hat{\bf I}_i\cdot \hat{\mathbf r}_{ij}\big)\!\big(\hat{\bf I}_j\cdot \hat{\mathbf r}_{ij}\big) \Big]\text{,}
\end{align}
where $g_{ij} = \hbar \mu_0 \gamma_{i} \gamma_{j}/4 \pi |\mathbf{r}_{ij}|^3$ is the coupling strength, 
$\hat{\bf I}_i = (\hat{I}_i^x, \hat{I}_i^y, \hat{I}_i^z)$ is the vector containing spin operators for each of the spatial dimensions, and $\mathbf{r}_{ij}=\mathbf{r}_i-\mathbf{r}_j$  is the vector joining the two interacting spins with  $\hat{\bf r}_{ij} = (\hat{x}_{ij},\hat{y}_{ij},\hat{z}_{ij})$ being its corresponding unit vector. Here, $\gamma_i$ is the gyromagnetic ratio of nucleus at site $i$, and $\mu_0$ is the permeability of the vacuum.

One can express the interactions in terms of raising and lowering spin operators $\hat{I}^\pm_i = \hat{I}^x_i \pm i\hat{I}^y_i$.
In an interaction picture with respect to the self-energy terms of the system Hamiltonian, $\hbar \omega_i \hat I^z_i$, with $\omega_i = \gamma_i B$ with $B$ being a homogenous magnetic field pointing in the $z$-direction, perpendicular to the two-dimensional plane of hBN (parallel to the hexagonal c axis), the interaction term acquires a time dependence according to 
\begin{align}
\hat{I}_i^\pm \rightarrow \hat{I}_i^\pm e^{\pm i \omega_i t}, \qquad \hat{I}_i^z \rightarrow \hat{I}_i^z \text{.}
\end{align}
Under the rotating wave approximation, one can neglect the effect of all time-dependent terms in Eq.~(\ref{eq:dipHam}), reducing the spin-spin interaction to
\begin{align}\label{a6}
\hat H_{\rm dd}^{ij} &= g_{ij}\Big[ \hat{I}_i^z \hat{I}_j^z -\frac{1}{4}\big(\hat{I}_i^+ \hat{I}_j^- + \hat{I}_i^- \hat{I}_j^+\big)\Big], 
\end{align}
 for spins with comparable gyromagnetic ratioes like nuclear spins---such that in the interaction picture terms $I_i^+I_j^-$ rotate slowly compared to $g_{ij}$. 

For the interaction between the defect electron spin with the surrounding nuclear spins, which have gyromagnetic ratios differing by up to 3 orders of magnitude, Eq.~(\ref{hamdd}) approximately becomes
\begin{align}
 \hat H^{{\rm e}i}_{\rm dd}&\approx g_{{\rm e}i} \hat{S}^{z}\:\hat{I}^z_{i}\text{ , }
\end{align}
where $g_{{\rm e}i}= \hbar \mu_0 \gamma_{e} \gamma_{i}/4 \pi |\mathbf{r}_{i}|^3$  is the hyperfine interaction constant of the electron spin to the nuclear spin at
the $i$th lattice site with $\mathbf{r}_i$ being the displacement vector pointing from electron to the $i$th nuclear.
In practical applications and for high magnetic field regime that we are considering in this work, the defect can be modeled as a two-level system described by the Pauli matrices $\hat{\sigma}^i$ (for $i=x,y,z$), whose dynamics can be accessed and controlled in the lab via microwave and optical fields. Consequently, we will refer to the defect as a qubit for the purposes of the following discussion, without loss of generality.
 
Finally, putting everything together the Hamiltonian of our interest is 
\begin{align}
\hat H &= \dfrac{\hbar}{2}\,\omega \hat \sigma^z +\hbar\sum_i \omega_{i} \hat I_i^z +\half\sum_{i} g_{{\rm e}i} \hat{\sigma}^{z}\:\hat{I}^z_{i} \nonumber\\
&+\sum_{i<j} g_{ij}\Big[ \hat{I}_i^z \hat{I}_j^z -\frac{1}{4}\big(\hat{I}_i^+ \hat{I}_j^- + \hat{I}_i^- \hat{I}_j^+\big)\Big] ,
\label{ham}
\end{align}
where $\hbar\,\omega\equiv D-\gamma_e B$ is the energy splitting of the two lowest levels of the defect electron spin, while the sums run over all the nuclear spins with their corresponding spin operators, which will be different for each kind of spin. The second line describes the dipolar coupling between the nuclear spins.

The unitary time evolution operator of the system using Hamiltonian (\ref{ham}) is given by
\begin{equation}
\hat{U}(t)=\exp\!\big(\!-i\hat{H}t/\hbar\big).
\end{equation}
The initial state of the electron spin is set in a maximally coherent superposition state, an eigenstate of the operator $\hat{\sigma}^x$, while the surrounding nuclear spins are considered in their individual thermal state at a given temperature. Therefore the density matrix of the joint system is
\begin{equation}
\rho_0 = \frac{1}{2}\left(\begin{matrix} 1&1\\1&1\\\end{matrix}\right)\otimes\bigotimes_i\varrho_{i},
\end{equation}
where $\varrho_i$ are thermal states of 
the nuclear spin systems. 
Therefore, the time evolved density matrix of the joint system can be found by
\begin{equation}
\rho(t)=\hat{U}(t)\rho_0\hat{U}^\dagger(t).
\end{equation}

Under pure dephasing, the reduced density matrix of the central spin evolves as
\begin{equation}
\rho_{\rm S}(t)=\dfrac{1}{2}\left(\begin{array}{cccccc}
1 &e^{-\Gamma_{\rm sb}(t)}\\
e^{-\Gamma_{\rm sb}(t)} & 1 \\
\end{array}\right)\text{ , }
\end{equation}
where $ \Gamma_{\rm sb}(t) $ indicates the decoherence function due to nuclear spin bath~\cite{jafari2024}. 
Consequently, we can estimate the coherence time in the presence of the nuclear spin bath, $T'_2$, from the envelope of oscillations of $\mean{\hat{\sigma}^x (t)}$, when it reaches the value of~$ \half\vert\expval{\hat{\sigma}^x(0)}\vert$,
where $\mean{\hat{\sigma}^x (0)}=1$. 
The expectation value of the spin operator $\hat{\sigma}^x$
can be obtained using
\begin{equation}
\mean{\hat{\sigma}^x(t)}= \text{Tr}(\rho_{\rm S}(t)\hat{\sigma}^x)= \exp[-\Gamma_{sb}(t)]. 
\end{equation}

To estimate $\mean{\hat{\sigma}^x(t)}$, the maximum number of nuclear sites that can be taken into consideration is largely limited by the computational cost.  
Given the available computational resources, this number cannot exceed 10 without significantly increasing the computation time. Therefore, as described in the following, we use a combined strategy to obtain a reliable coherence time estimation.

We consider the mean-field approximation in which the dynamics of the spin bath and the \vnnb\ spin are treated independently. Therefore, the dynamics of \vnnb\ is integrated exactly under a time-dependent Hamiltonian that incorporates the expectation value of the bath operators over time
\begin{equation}
\label{eq:NVHam}
\hat{H}_{\rm S}(t) = \half\Big[\hbar\,\omega + \sum_{i=1}^N g_{{\rm e}i} \big\langle\hat{I}^z_i(t)\big\rangle\Big]\hat{\sigma}^z.
\end{equation}
Let us introduce $\Omega(t)\equiv \half(\hbar\,\omega +\sum_i g_{{\rm e}i} \mean{\hat{I}_i^z(t)})$ as the effective time-dependent level splitting of the electron spin in \vnnb. The spin bath Hamiltoinian is also given by
\begin{align}
\hat{H}_{\rm B}&=  \hbar\sum_i \omega_{i} \hat I_i^z +\sum_{i<j} g_{ij}\Big[ \hat{I}_i^z \hat{I}_j^z -\frac{1}{4}\big(\hat{I}_i^+ \hat{I}_j^- + \hat{I}_i^- \hat{I}_j^+\big)\Big].
\label{bath}
\end{align}
To simulate the dynamics of the spin bath, we will turn to numerical simulations based on the Gaussian state formalism, which enables us to investigate larger spin systems.

Note that the Hamiltonian in Eq.~\eqref{eq:NVHam} represents a central spin coupled to a spin bath, influenced by a time-dependent noisy magnetic field~\cite{jafari2024}. Our primary focus is to understand how this noise impacts the dephasing time of the central spin.

\subsection{The hybrid Gaussian method}
It is well-established that bosonic fields initially in Gaussian states maintain their Gaussian character when evolving under Hamiltonians quadratic in creation and annihilation operators. 
As Gaussian states are fully characterized by their first and second-order moments, deriving the equations of motion for these moments suffices to describe the system's dynamics. Notably, these equations have a dimensionality that increases linearly with the number of modes, making simulation feasible.

To model the behavior of large spin baths, we utilize the Holstein-Primakoff transformation (HPT) and its approximation (HPA) to compute the time evolution of 
spins represented by bosonic states~\cite{Holstein1940}. While the Hamiltonian of the nuclear spin bath can be mapped to bosons using the HPT, the resulting bosonic Hamiltonian is not quadratic.
Thus, to simplify the computation, we adopt a linear approximation—the lowest order of the HPA—which is most effective for spin states near their ground state.

In this approximation, the spin operators of a spin-s particle are transformed as follows~\cite{Holstein1940}
\begin{align}
\hat{I}^z &= \hat{a}^\dag \hat{a} - s \one \\
\hat{I}^- &= \sqrt{2s - \hat{a}^\dag \hat{a}}~\hat{a} \approx \sqrt{2s}~\hat{a} \\
\hat{I}^+&= \hat{a}^\dag \sqrt{2s - \hat{a}^\dag \hat{a}} \approx \hat{a}^\dag \sqrt{2s},
\label{eq:HPz}
\end{align}
where $\hat{a}^\dag$ ($\hat{a}$) is the bosonic creation (annihilation) operator with commutator $[\hat{a},\hatd{a}]=\one$.

Under this mapping the Hamiltonian of the spin bath, Eq.~\eqref{bath}, takes the form
\begin{equation}
\label{eq:GaussHam}
\hat{H}_{\rm Gauss} (t)= \sum_{i=1}^N \tilde \omega_i (t) \hat{a}^\dag_i \hat{a}_i - \sum_{ i<j}^N B_{ij} \big(\hat{a}_i \hat{a}_j^\dag  + \hat{a}_i^\dag \hat{a}_j \big),
\end{equation}
where $N$ is the number of nuclei and we have introduced $\tilde \omega_i (t)= \hbar\,\omega_i  + \half\sum_{ j \ne i}^N g_{ij}\big(\overline{n}_i(t) -2s_j \big)$, and $B_{ij} = \frac{1}{2}g_{ij}\sqrt{s_i s_j}$. Here, $\overline{n}_i = \langle \hat{a}_i^\dag \hat{a}_i \rangle$ is the instant occupation number of the $i$th spin site. 
Notice that in order to get a quadratic Hamiltonian we have made a mean-field type approximation $\hat{a}^\dag_i\hat{a}_i \hat{a}_j^\dag \hat{a}_j \rightarrow \langle\hat{a}^\dag_i\hat{a}_i\rangle \hat{a}_j^\dag \hat{a}_j $, where the expectation value is taken over the instantaneous state of the system. 

The Hamiltonian presented above can be expressed in a more compact form as $\hat{H}_{\rm Gauss} (t) = \hat{\bf R}^{\dagger} V(t)\hat{\bf R}$.  In this notation, we define the bosonic operator vector  $\hat{\bf R}=( \hat{a}_{1},\hat{a}_{2}, \cdots, \hat{a}_{N})^{\sf T}$ and $V(t)$
is an $N\times N$ matrix.

As we assume that the initial state of the spin bath is thermal, it follows that $\mean{\hat{\bf R}}=0$. This expectation value remains zero due to the purely quadratic form of Eq.~\eqref{eq:GaussHam}. Beginning with the von Neumann equation, it can be easily demonstrated that the evolution of the covariance matrix  $\Gamma_{i,j} = \langle \hat{R}_i^\dag \hat{R}_j\rangle$ is described by the following expression
\begin{equation}\label{a18}
\dot \Gamma = - i \,[V(t), \Gamma] \text{ , }
\end{equation}
with matrix $V(t)$ introduced such that $V_{ii} = \hbar \tilde \omega_i(t)$ and the off-diagonal elements given by
\begin{equation}
V_{ij} = B_{ij} \qquad {\rm for } \  i \ne j.
\end{equation}
The formal solution of Eq.~\eqref{a18} is
\begin{equation}
\Gamma(t) =\mathcal{U}(t)\,\Gamma(0) \,\mathcal{U^\dag}(t)\text{ , }
\label{i10}
\end{equation}
where $\mathcal{U}(t)=\exp\left[ -i\int_0^t V(s)ds\right] $ and $\Gamma(0)$ is the initial covariance matrix. However due to excitation preserving nature of the Hamiltonian in Eq.~\eqref{eq:GaussHam} the system dynamics of such global thermal state remains trivial and does not give the true evolution of the coherence of the central spin, see e.g. Ref.~\cite{SHAYEGANI2024129966}. Therefore, to estimate the impact of the spin bath on the coherence time of the defect electron spin using the HP approximation, we employ the randomized bath method in our computations, which effectively approximates the behavior of the spin bath.


Our simulation approach consists of several key steps. First, the spin bath is prepared with all spins in their ground state. Next, spin excitations are introduced in the bath randomly with a probability given by the Boltzmann distribution that depends on the bath temperature. In other words, the state of each spin is treated as a local thermal state whose occupation number $\overline{n}_i$ equals the  excitation number which in turn is assigned through a Boltzmann weighted random number. The \vnnb\ defect is then initialized in an eigenstate of the operator $\hat{\sigma}^x$. 
Following this, the covariance matrix of the spin bath is time evolved using Eq.~\eqref{i10}. The \vnnb\ defect undergoes time evolution according to the Hamiltonian described in Eq.~\eqref{eq:NVHam}. We measure the expectation value of the operator $\hat{\sigma}^x$  at every value of $t$. This entire procedure is repeated many times, with a new random sample of the spin bath used for each iteration. Finally, we compute the average value of $\expval{\hat{\sigma}^x(t)}$ over all random samples of the spin bath and estimate $T'_2$ from the envelope of  $\vert\!\expval{\hat{\sigma}^x }\!\vert$.

\begin{figure}[t]
\includegraphics[width=1\columnwidth]{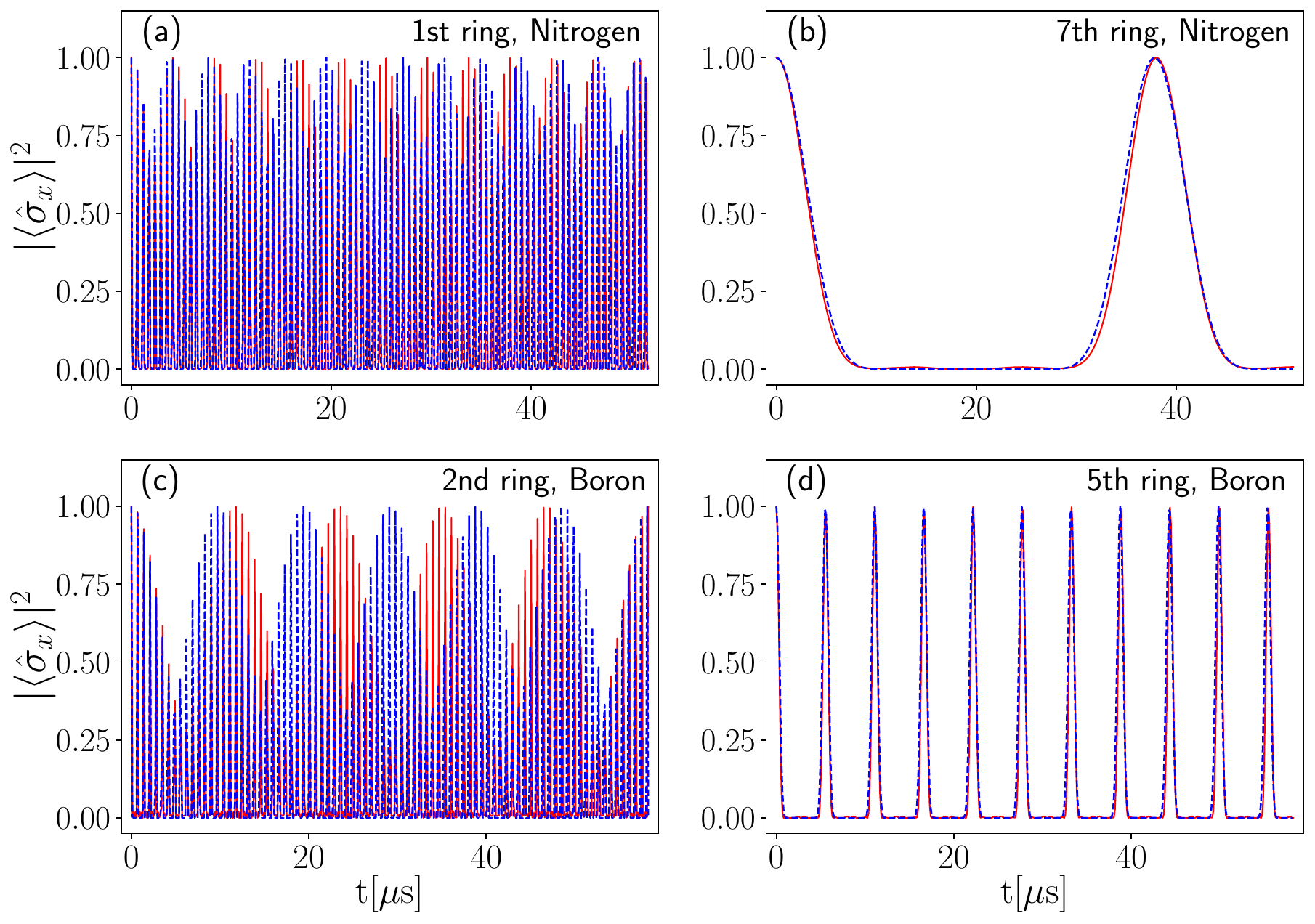}\\
\caption{%
Comparison of the exact numerical simulation and HPA method for the time evolution of the expectation value of electron spin operator $\hat \sigma_x$ in the presence of 3 nitrogens and 3 borons in the different rings and for the spin bath temperature $\mathbf{T}=10^{-1}\:\mathrm{K} $. 
We used 100 samples and $B = 1\:\mathsf{T}$. The blue and red colors represent simulations based on the equations \eqref{ham} and \eqref{eq:NVHam}, corresponding to the exact and the HPA approach, respectively.
}
\label{fig1}%
\end{figure}
%
%

%
\begin{figure}[t]
\includegraphics[width=\columnwidth]{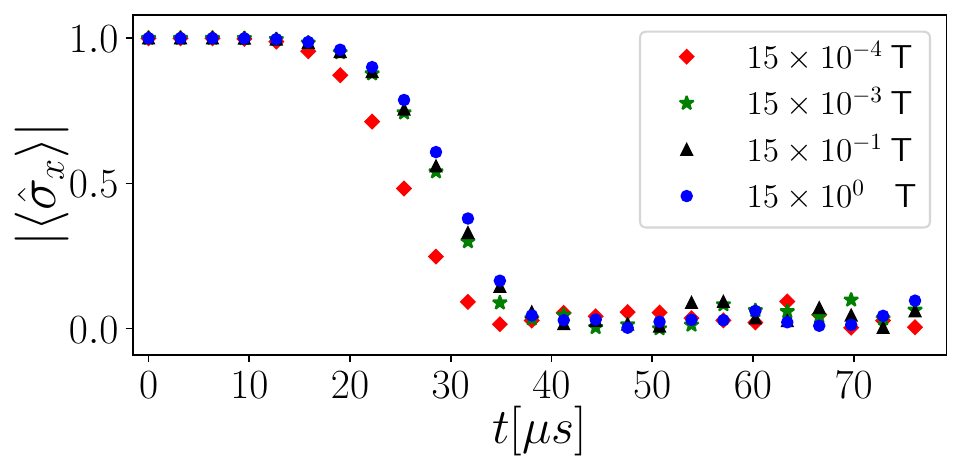}
\caption{%
The plot of the electron expectation value of $\sigma_x$ in the presence of a lattice composing of 30 spins (18 borons and 12 nitrogens ) for different values of external magnetic field. A spin echo pulse has been used along with 200 samples and $ \mathbf{T}=10^{-1}\: K$
}
\label{fig2}
\end{figure}
\begin{figure}[b]
\includegraphics[width=\columnwidth]{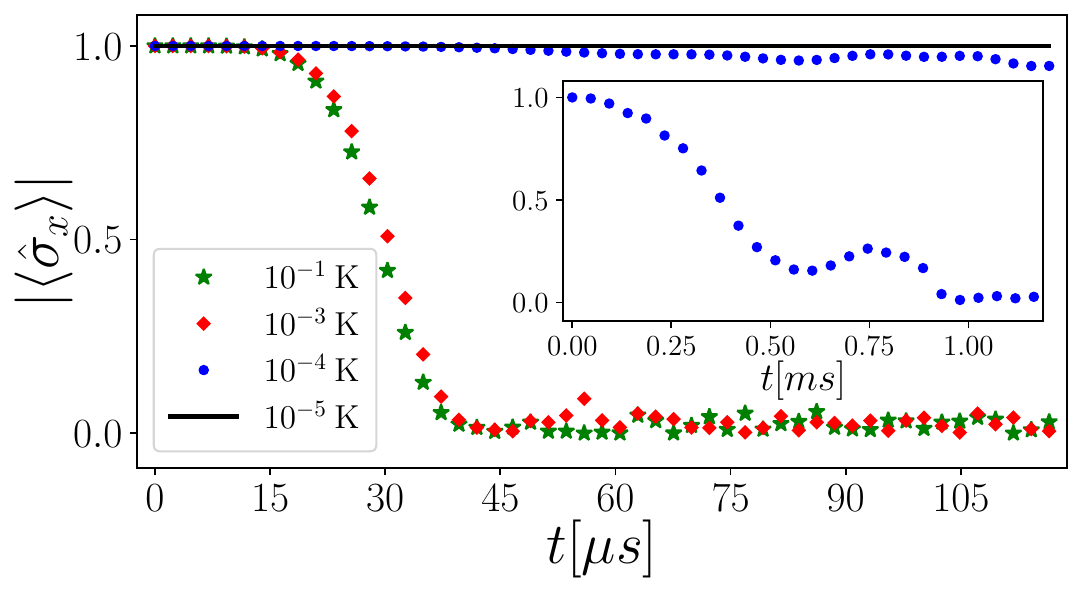}
\caption{%
The Coherence dynamics of the \vnnb\ defect, $ |\langle \hat \sigma_x\rangle| $, are shown in the presence of a lattice consisting of 240 nuclei spins (120 borons and 120 nitrogens ) and  for different values of the spin bath temperature where a spin echo pulse
 is applied. The inset gives the coherence dynamics for $ \mathbf{T}=10^{-4}\:\mathrm{K}$. The other parameter is  $B=1\:\mathsf{T}$ with 800 samples.
}
\label{fig3}
\end{figure}
\subsection{Results}
In order to obtain the decoherence time, $T'_2$, a Hahn echo pulse sequence is required, where the spin is initially prepared in a superposition state and then it is allowed to evolve freely for time $t/2$. A $\pi$-pulse is then applied and after another free evolution time interval $t/2$ the echo is measured~\cite{Hahn}. Moreover, we define ‘rings’ around the defect where the nuclei placed  in equal distances from \vnnb.
The first ring 
consists of 3 nearest neighbor nitrogens,  6 borons in the next nearest neighbor are  situated in the second ring, 3 nitrogens lie in the third ring, 6 nitrogens in the fourth ring, 6 borons in the fifth ring, 6 borons in the sixth ring, and so on. 

Before studying large-size systems, we first compare the results of the exact solution and the bosonic approximation for a few nuclei. This allows us to benchmark the performance of our approximate Gaussian method.  We make reference to the numerical outcomes of equations \eqref{ham} and \eqref{eq:NVHam} as exact and HPA approaches, respectively.

The time evolution of electron coherence, the off-diagonal elements of the electron reduced density matrix in a uniform background magnetic field of unit Tesla are shown in Fig.~\ref{fig1}. We consider three nitrogen nuclei in the first ring and the 7th ring in Fig.~\ref{fig1}(a) and (b), respectively, as well as three boron nuclei in the second ring and 5th ring in Fig.~\ref{fig1}(c) and (d), respectively.  As can be seen, the exact numerical simulation is in good agreement with the  bosonic numerical approach, where 100 samples have been used and the temperature is 
$\mathbf{T}=10^{-1}$~K. We also note that the exact and the approximate bosonic simulations match very well when the nuclei are farther away from the defect. In addition, we have investigated a long time behavior  of $ \langle \hat \sigma_x\rangle $ for 3  nearest nitrogens and we find that  there is a periodic behavior  and the coherence revives at later times. It means that  only 3 nitrogens can not led to defect decoherence and other nuclei play an important role. 

In the following, we take into account the coherence time in the presence of 30 nuclei, 18 borons and 12 nitrogens, in which six rings are chosen.  Here, the temperature is assumed $ \mathbf{T}=10^{-1}$~K at which the spin bath becomes completely thermal and the number of samples is 200.  Figure ~\ref{fig2} demonstrates  $T'_2\approx 30 \:\mu s$ and it implies that different magnetic fields do not have a significant effect on the coherence time. Red diamond shows coherence dynamic for $B= 15 \times 10^{-4}\:\mathsf{T}$.

The defect coherence dynamics in the presence of a lattice consisting of 240 nuclei, 120 borons and 120 nitrogens,  are displayed in Fig~\ref{fig3}. As can be observed, for high temperatures,  its behavior is very similar to Fig.~\ref{fig2}. It indicates that $T'_2$ of \vnnb\ is approximately $ 30~\mu s$ and implies that $T'_2$  does not apparently change by increasing the number of nuclei. It means that the nearest nuclear spins to the defect play a significant role in the coherence time. In addition, we find for temperatures greater than millikelvin, the coherence time does not vary and the spin bath decoherence function is obtained through curve fitting, expressed as $\Gamma_{\rm sb}(t)=(
{0.92\, t}/{T'_2})^6$.
However, for temperatures less than millikelvin, for example, $\mathbf{T}=10^{-4}$~K, the coherence time is at the order of millisecond and by further decreasing the temperature 
we do not observe the decoherence.  In fact, the revival is full for very low temperatures such as $\mathbf{T}=10^{-5}$~K, which is practically the zero temperature for both boron and nitrogen spin at an external magnetic field of $B=1\:\mathsf{T}$. Note that we have obtained the inhomogeneous dephasing time as $T_2^*\sim 40\: ns$ by removing Hahn spin echo. 

In the following section, we investigate the effect of spin-phonon interaction on the coherence time by employing the Debye approximation and the master equation approach.

\section{The phononic bath}\label{phbath}%
In the previous section, we examined the dephasing time of the \vnnb\ center resulting from hyperfine interactions with the nuclear spin bath. In this section, we shift our focus to a single \vnnb\ center interacting with hBN lattice phonons. We assume that the decoherence effects arising from the lattice phonons do not interfere with the spin bath interactions discussed earlier.
 
To model the spin-phonon dephasing dynamics, we consider the following Hamiltonian
\begin{align}
\hat{H} =\hat{ H}_{\rm s}+\hat{H}_{\rm s-ph}+\hat{H}_{\rm ph},
\end{align}
where the first term, $\hat{ H}_{\rm s}$, represents the ground state spin Hamiltonian of the \vnnb\ center; the second term, $\hat{H}_{\rm s-ph}$, describes the interaction between the spin state and lattice phonons; and the third term, $\hat{H}_{\rm ph}$, accounts for the phonon bath.

In the presence of a static magnetic field $B_0$ oriented along the z axis, the spin Hamiltonian for the \vnnb\ center can be expressed as
\begin{equation}\label{eq21}
\hat{H}_{\rm s}= \hbar\, D \,(\hat{S}^{z})^2+\hbar\,\omega \,\hat{S}^{z},
\end{equation}
The phonon Hamiltonian is given by
\begin{equation}
\hat{H}_{\rm ph}=\sum_k \hbar\,\omega_k \,b^\dag_k b_k,
\end{equation}
where $\omega_k$ is the frequency of the $k$-th mode and  $b_k$
and $b^\dag_k$ are the boson annihilation and
creation operators, respectively,satisfying $[b_k, b^\dag_{k'}]=\delta_{k,k'}$.

For pure dephasing dynamics in a system with spin 
$S = 1$, the spin-phonon Hamiltonian is described by the following equation~\cite{Maze}
\begin{equation}
\hat{H}_{\rm s-ph}= \hat{E}_{\rm z} (\hat{S}^{z})^2 ,
\end{equation}
where $\hat{E}_{\rm z}$ is 
\begin{align}
\hat{E}_{\rm z}= \sum_{k}\hbar\,\lambda_k (b_k + b^\dag_k) 
+\sum_{k, k'}\hbar\,\lambda_{k\,k'}(b_k+b^\dag_k)(b_{k'}+b^\dag_{k'}),
\end{align}
where $\lambda_k$ and $\lambda_{k\,k'}$ are the linear and quadratic spin-phonon coupling constants, respectively. 

\begin{figure}[tb]
\includegraphics[width=0.9\columnwidth]{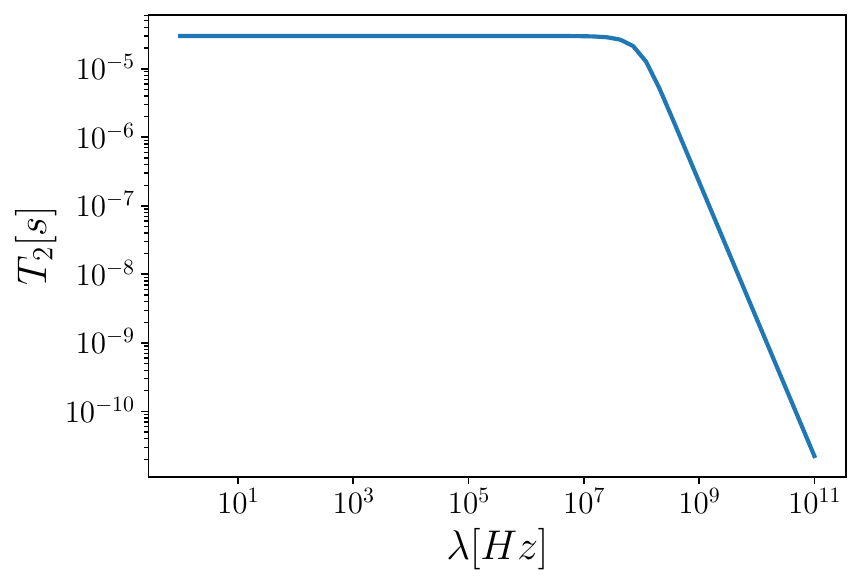}\\
\caption{%
The coherence time, 
$T_{2}$, with respect to the two-phonon coupling constant, $\lambda=\lambda_{00}$. 
}
\label{fig4}%
\end{figure}
The master equation governing the dynamics of the spin defect can be obtained as (for more details, see Appendix)
\begin{align}\label{eq26}
\dot{\rho}_s(t)=-\gamma\left((S^z)^2 \rho_s (t)+\rho_s (t)(S^z)^2 -2(S^z)^2 \rho_s (t)(S^z)^2\right),
\end{align}
where the decay rate, $\gamma$, is given by 
\begin{align}
\gamma = 2\pi \dfrac{8\pi^2 A^2 \lambda^2_{00} }{\nu^4_{s}(\omega_D)^{4\upsilon}}\left( \frac{k_B T}{\hbar}\right)^{4\upsilon+3}\! (4\upsilon+2)! ,
\end{align}
for $\hbar\,\omega\gg k_B T$ 
and
\begin{align}
\gamma = 2\pi \dfrac{8\pi^2 A^2 \,\lambda^2_{00}\, \omega_D }{\nu^4_{s}(4\upsilon+1)}\left( \frac{k_B T}{\hbar}\right) ^{2},
\end{align}
for $\hbar\,\omega\ll k_B T$. In the decay rate, the Debye frequency is denoted by $\omega_D$, while 
$\upsilon $ represents a phenomenological parameter that characterizes the relationship between the spin-phonon coupling strength and the phonon frequency. Additionally, the factor $A$ refers to the area of a unit cell, 
$\nu_{s}$ is the speed of sound in the hBN lattice, and $\lambda_{00}$ indicates the two-phonon coupling constant for acoustic phonons at the Debye frequency.The above master equation indicates that single-phonon transitions do not impact the coherence time. By considering the two lowest defect electron spin levels, the decoherence function associated with the phononic bath can be expressed as $\Gamma_{pb}(t)=\gamma\,t$.

Now, we take into account the effect of spin and the phonon baths together. These two baths are considered independently, allowing us to express the total decoherence function as follows:
\begin{align}
F(t) &= \exp\!\big[-\Gamma_{\rm tot}\, (t)\big]= \exp\!\big[-\Gamma_{\rm pb}(t)-\Gamma_{\rm sb}(t)\big]\nonumber\\
&=\exp\!\left[ - \gamma\, t - \Big(\dfrac{0.92\, t}{T'_2}\Big)^6 \right].
\end{align}
The first term in the decoherence function arises from the phonon bath, while the second term corresponds to the spin bath. Consequently, the coherence time at the room temperature can be determined from the decoherence function, achieving the condition $F(T_2)=1/2$.

The behavior of the coherence time with respect to the two-phonon coupling constant is shown in Fig.~\ref{fig4}. The hBN Debye energy is approximately $\hbar \omega_D\approx 175 ~meV$ ~\cite{rost}, and the speed of sound in hBN is assumed to be $\nu_{s}\approx 10^4~ m/s$. The phenomenological parameter is taken as $\upsilon=3/8$ based on Ref.~\cite{Gottscholl2021}, and $A= 3 \,a_0 \sin(\pi/6)$, where $a_0\approx 1.5~ \AA =1.5\times 10^{-10}~m $  is the hBN lattice constant. As shown in Fig.~\ref{fig4}, the phonon bath can reduce the coherence time when the coupling constant exceeds $\lambda_{00}>10^8~ {\rm Hz}$, which leads to a dephasing time in the order of nanoseconds. Specifically, $T_2\approx 30~\mu s $ for $ 0\leqslant \lambda_{00} < 10^8~ {\rm Hz}$.

\section{CONCLUSION}\label{con}%
In this work, we have presented a theoretical description for understanding the coherene time of \vnnb\ in  a mono-layer of the hBN lattice. We have considered the interaction between the spin of the \vnnb\ center and the nuclear spin bath of the surrounding nuclei, as well as the effects of thermal vibrations within the hBN lattice. To simulation the coherence time of the \vnnb\, we have employed the Holstein-Primakoff transformation and the quantum master equation approach, utilizing the Debye approximation to account for lattice vibrations.

Our results indicate that the primary source of the \vnnb\ pure dephasing time arises from interactions with the spin bath when the two-phonon coupling constant is less than 
$10^8~{\rm Hz}$. Additionally, we observed that the single-phonon transitions do not have an impact on the coherence time. In contrast, two-phonon transitions can play a critical role, particularly when the coupling constant is large. 

Our analysis contributes to a deeper understanding of the factors that influence the coherence properties of color centers in two-dimensional materials. However, it is important to recognize that other factors likely affect qubit decoherence. For instance, interactions with electric fields generated by charges within the solid, as well as the coupling between the center’s electrons and crystal strain fields, can lead to dephasing. Future research should explore these additional sources of decoherence, extending the focus beyond magnetic noise and incorporating lattice vibrations without relying on the Debye approximation.

%
%
\begin{acknowledgements}
FTT would like to thank J. S. Pedernales and R. Asgari for useful discussions. 
\end{acknowledgements}
\appendix*

\section{Derivation of the master equation (\ref{eq26})}\label{app}
To obtain the spin reduced master equation, Eq.(\ref{eq26}), we consider
the following total Hamiltonian 
\begin{align}\label{eq22}
\hat{H}=\hat{H}_{\rm s}+\hat{H}_{\rm p}+\hat{H}_{\rm s-ph},
\end{align}
where $\hat{H}_{\rm s}$ represents the spin Hamiltonian for the \vnnb\ center, and 
$\hat{H}_{\rm p}$ is the phonon Hamiltonian, defined as
\begin{equation}
\hat{H}_{\rm s}= \hbar\, D \,(\hat{S}^{z})^2+\hbar\,\omega \,\hat{S}^{z},
\end{equation}
\begin{equation}
\hat{H}_{\rm ph}=\sum_k \hbar\,\omega_k(b^\dag_k b_k).
\end{equation}
The spin-phonon interaction is characterized by~\cite{Maze}
\begin{equation}
\hat{H}_{\rm s-ph}= \hat{E}_{\rm z} (\hat{S}^{z})^2 ,
\end{equation}
where $\hat{E}_{\rm z}$ is associated with the irreducible representation $A'_1$~\cite{Doherty2013, Abdi2018}, and is defined as
\begin{align}
\hat{E}_{\rm z} &= \sum_{k \in A'_1}\hbar\,\lambda_k (b_k + b^\dag_k)\nonumber\\ 
&+\sum_{k\otimes k' \in A'_1}\hbar\,\lambda_{k\,k'}(b_k+b^\dag_k)(b_{k'}+b^\dag_{k'}),
\end{align}
where $\lambda_k$ and $\lambda_{k\,k'}$ are the linear and quadratic spin-phonon coupling constants, respectively. 

In the following, we rewrite the total Hamiltonian, Eq.(\ref{eq22}),
by adding and subtracting the expectation value of $\mean{\hat{H}_{\rm s-ph}}$ to the bath operators as
\begin{align}
\hat{H}=\hat{H}'_{\rm s}+\hat{H}_{\rm p}+\hat{H}'_{\rm s-ph},
\end{align}
where
\begin{align}
\hat{H}'_{\rm s}= \hat{H}_{\rm s}+ \mean{\hat{H}_{\rm s-ph}},
\end{align}
and
\begin{align}
\hat{H}'_{\rm s-ph}=\hat{H}_{\rm s-ph}-\mean{\hat{H}_{\rm s-ph}}.
\end{align}
The expectation value of the spin-phonon coupling to the bath operators at equilibrium is as following
\begin{align}
\mean{\hat{H}_{\rm s-ph}}&=\mean {\hat{E}_{\rm z}} (S^z)^2\nonumber\\
&=\sum_{k\otimes k'\in A'_1}\!\! \hbar\lambda_{k\,k'}(2 \,\bar{n}_k+1)\delta_{k,k'}(S^z)^2,
\end{align}
where it is obtained from 
the non-vanishing
expectation value of the quadratic spin-phonon coupling and $ \bar{n}_k= [\exp(\hbar\,\omega_k/k_B\, T)-1]^{-1}$ is  the average occupation number of the phonon mode 
$ k $ at temperature $ T $.

Thus, the redefined Hamiltonians can be written as
\begin{align}
\hat{H}'_{\rm s}= \bigg(\!\hbar D+\!\!\!\sum_{k\otimes k'\in A_1}\!\!\!\hbar \lambda_{k\,k'}(2 \bar{n}_k+1)\delta_{k,k'}\!\bigg)\!\big(\hat{S}^z\big)^2+ \hbar\,\omega \hat{S}^z,
\end{align}
and
\begin{align}
&\hat{H}'_{\rm s-ph}= \bigg[  \sum_{k \in A'_1}\hbar\,\lambda_k (b_k+b^\dagger_k)+\\
&\sum_{k\otimes k'\in A'_1}\!\!\!\! \hbar\lambda_{k\,k'}\!\left(\! (b_k+b^\dagger_k)
 (b_{k'}+b^\dagger_{k'})-(2 \bar{n}_k+1)\delta_{k,k'}\!\right)\!\bigg]\! (\hat{S}^z)^2.\nonumber
\end{align}

As the spin Hamiltonian commutes with the spin-phonon interaction Hamiltonian, 
 $[\hat{H}'_{\rm s} , \hat{H}'_{\rm s-ph}]=0$, then the spin-phonon Hamiltonian in the interaction picture can be expressed as
\begin{align}
\hat{H}'_{\rm s-ph}(t)=& e^{i \hat{H}_ {\rm p} t/\hbar} \hat{H}'_{\rm s-ph} e^{- i \hat{H}_{\rm p} t/\hbar}\nonumber\\
=& \bigg[\sum_{k\in A'_1}\hbar\,\lambda_k \Big(b_k e^{i \omega_k t}+b^\dagger_k e^{-i \omega_k t}\Big)\\
&+\sum_{k\otimes k'\in A'_1}\hbar\, \lambda_{k\,k'}\Big( b_k b_{k'} e^{i  t (\omega_k +\omega_{k'})}\nonumber\\
&+b^\dagger_k b_{k'} e^{-i t (\omega_k -\omega_{k'})}+ b_k b^\dagger_{k'} e^{i t (\omega_k -\omega_{k'})}\nonumber\\
&+b^\dagger_k  b^\dagger_{k'} e^{- i t (\omega_k +\omega_{k'})}-(2 \bar{n}_k+1)\delta_{k,k'}\Big) \bigg](\hat{S}^z)^2.\nonumber
\end{align}
Using the rotating wave approximation, the interaction Hamiltonian 
can be simplified by neglecting fast oscillating terms and
can be written as 
\begin{align}
\hat{H}'_{\rm s-ph}(t)=& \bigg[\sum_{k\in A'_1}\hbar\,\lambda_k \Big(b_k e^{i \omega_k t}+b^\dagger_k e^{-i \omega_k t}\Big)\\
&+\!\!\!\sum_{k\otimes k'\in A'_1}\!\!\!\hbar\,\lambda_{k\,k'}\Big(b^\dagger_k b_{k'} e^{-i t (\omega_k -\omega_{k'})}\nonumber\\
&+ b_k b^\dagger_{k'} e^{i t (\omega_k -\omega_{k'})}-(2 \bar{n}_k+1)\delta_{k,k'}\Big)\bigg](\hat{S}^z)^2.\nonumber
\end{align}
We rewrite the above relation as 
\begin{align}
\hat{H}'_{\rm s-ph}= [A(t)+B(t)](\hat{S}^z)^2,
\end{align}
where
\begin{align}
A(t)=&\sum_{k\in A'_1} \hbar\,\lambda_k \Big(b_k e^{i \omega_k t}+b^\dagger_k e^{-i \omega_k t}\Big),\\
B(t)=&\!\!\!\sum_{k\otimes k'\in A'_1}\!\!\!\hbar\,\lambda_{k\,k'}\Big(b^\dagger_k b_{k'} e^{-i t (\omega_k -\omega_{k'})}+ b_k b^\dagger_{k'} e^{i t (\omega_k -\omega_{k'})}\nonumber\\
&\quad-(2 \bar{n}_k+1)\delta_{k,k'}\Big).
\end{align}
Up to second order in the spin-phonon coupling, the spin reduced density matrix, $\rho_s(t)$, evolves
according to the quantum master equation
\begin{align}  
	\begin{split}
\dot{\rho}_s(t) =& -\frac{1}{\hbar^2} \!\int_0^{\infty}\! \tr_{ph} \left[ H'_{s-ph}(t), \right. \\
& \left. \big[H'_{s-ph}(t-t'), \rho_s (t) \otimes \rho_{ph}\big] \right] dt'.  
	\end{split}
\end{align} 
Substituting the expression for the spin-phonon coupling Hamiltonian $\hat{H}'_{\rm s-ph}$ into the above equation, 
we obtain
\begin{align}
\dot{\rho}_s(t)=-\frac{1}{\hbar^2}\!&\int_0^{\infty}\! dt'\big\langle (A(t)+B(t))(A(t-t')+B(t-t'))\big\rangle\nonumber\\
&\left((S^z)^2 \rho_s (t)-(S^z)^2 \rho_s (t)(S^z)^2\right)\nonumber\\ 
-\frac{1}{\hbar^2}\!&\int_0^{\infty}\! dt'\big\langle (A(t-t')+B(t-t'))(A(t)+B(t))\big\rangle\nonumber\\
& \left( \rho_s (t)(S^z)^2-(S^z)^2 \rho_s (t)(S^z)^2\right).
\end{align}
The expectation values at equilibrium for linear interaction are given by
\begin{align}
\big\langle A(t)A(t-t')\big\rangle=\!\sum_{k\in A'_1}\hbar^2\lambda^2_k \left( (\bar{n}_k+1)e^{i\omega_k t'}+\bar{n}_k e^{-i\omega_k t'}\right),
\end{align}
while for quadratic interaction we have
\begin{align}
\big\langle B(t)B(t-t')\big\rangle=&2\!\!\!\sum_{k\otimes k'\in A'_1}\!\!\!\hbar^2\lambda^2_{k\,k'} \Big(\bar{n}_{k'}(\bar{n}_k+1)e^{i(\omega_k-\omega_{k'})t'} \nonumber\\
&+\bar{n}_k(\bar{n}_{k'}+1) e^{-i(\omega_k-\omega_{k'}) t'}\Big).
\end{align}
Additionally, we find that $\mean{B(t-t')A(t)}=\mean{A(t-t')B(t)}=0$.


Applying the relation $\int_0^{\infty}dt'e^{-i\varepsilon t'}=\pi \delta(\varepsilon)$, the master equation can be represented as 
\begin{align}
&\dot{\rho}_s(t)=-\pi \bigg( \sum_{k\in A'_1}\lambda^2_k \left( (\bar{n}_k+1)\delta(-\omega_k)+\bar{n}_k \delta(\omega_k )\right)\nonumber\\
&+2\sum_{k\otimes k'\in A'_1}\lambda^2_{k\,k'} \big(\bar{n}_{k'}(\bar{n}_k+1)\delta(-\omega_k+\omega_{k'})+ \\
&\bar{n}_k(\bar{n}_{k'}+1) \delta(\omega_k-\omega_{k'})\big)\!\bigg)\!\Big((S^z)^2 \rho_s (t)-(S^z)^2 \rho_s (t)(S^z)^2\Big)
\nonumber\\ 
&-\pi\bigg(\sum_{k\in A'_1}\lambda^2_k \left((\bar{n}_k+1)\delta(\omega_k)+\bar{n}_k \delta(-\omega_k)\right)
\nonumber\\
&+2\sum_{k\otimes k'\in A'_1}\lambda^2_{k\,k'} \big(\bar{n}_{k'}(\bar{n}_k+1)\delta(\omega_k-\omega_{k'})+ \nonumber\\
&\bar{n}_k(\bar{n}_{k'}+1) \delta(\omega_{k'}-\omega_k)\big)\!\bigg)\!\Big(\rho_s (t)(S^z)^2\! -(S^z)^2 \rho_s (t)(S^z)^2\Big).\nonumber
\end{align}
Assuming the Debye model is valid, i.e., with a linear dispersion relation given by $\omega_k=\nu_s k$, and  in the limit of continuous frequency, 
$\omega_k\rightarrow \omega$, one can
introduce the following scaling for the linear spin-phonon coupling constant
\begin{align}
\lambda_k\rightarrow \lambda(\omega)=\lambda_0\Big(\dfrac{\omega}{\omega_D}\Big)^\upsilon ,
\end{align}
for $0\leqslant \omega\leqslant\omega_D$. Here, $\lambda_0$ is the strength of the one-phonon coupling constant for acoustic
phonons at the Debye frequency $\omega_D$, and the parameter $\upsilon $ serves as a phenomenological parameter that characterizes the dependence of the spin-phonon coupling strength on the phonon frequency.

We introduce the density of states  (DOS) for acoustic phonons  in a two-dimensional lattice, based on the Debye approximation, as follows
\begin{align}
\sum_{k}\rightarrow A\int d\Omega \int k\,dk=2\pi A \int k\,dk=\dfrac{2\pi A }{\nu^2_{s}}\int^{\omega_D}_0\!\!\omega\, d\omega,
\end{align}
where $A$ is the area of a unit cell, and $\nu_{s}$ is the speed of sound in the hBN lattice.

In addition, we introduce the following scaling for
the quadratic spin-phonon coupling constant for the acoustic
phonon modes in the limit of continuous frequency
\begin{align}
\lambda_{k\,k'}\rightarrow  \lambda(\omega, \omega')=\lambda_{00}\Big(\frac{\omega}{\omega_D}\Big)^\upsilon \Big(\frac{\omega'}{\omega_D}\Big)^\upsilon,
\end{align}
where $\lambda_{00}$ is the two-phonon coupling constant for
acoustic phonons at the Debye frequency.
We also define 
\begin{align}
\sum_{k,k'}\rightarrow \dfrac{4\pi^2 A^2 }{\nu^4_{s}}\int^{\omega_D}_0\omega\, d\omega \int^{\omega_D}_0\omega' \,d\omega'.
\end{align}

Therefore, by regarding the following relation
\begin{align}
\sum_{k,k'}\lambda^2_{k\,k'}\rightarrow \dfrac{4\pi^2 A^2 \lambda^2_{00} }{\nu^4_{s}(\omega_D)^{4\upsilon}}\int^{\omega_D}_0 \omega^{2\upsilon+1} \, d\omega \int^{\omega_D}_0\omega'^{(2\upsilon+1)} \,d\omega',
\end{align}
we obtain
\begin{align}
&\sum_{k,k'}\lambda^2_{k,k'} \Big(\bar{n}_{k'}(\bar{n}_k+1)\delta(-\omega_k+\omega_{k'})\\
&\quad+\bar{n}_k(\bar{n}_{k'}+1) \delta(\omega_k-\omega_{k'})\Big)\nonumber\\
&=2\dfrac{4\pi^2 A^2 \lambda^2_{00} }{\nu^4_{s}(\omega_D)^{4\upsilon}}\int^{\omega_D}_0  \bar{n}(\omega)(\bar{n}(\omega)+1)\,\omega^{4\upsilon+2}\, d\omega \nonumber\\
&=\dfrac{8\pi^2 A^2 \lambda^2_{00} }{\nu^4_{s}(\omega_D)^{4\upsilon}}\int^{\omega_D}_0  \omega^{4\upsilon+2}
\dfrac{e^{\beta \omega}}{(e^{\beta \omega}-1)^2}
\,d\omega\nonumber\\
&=\dfrac{8\pi^2 A^2 \lambda^2_{00} }{\nu^4_{s}(\omega_D)^{4\upsilon}}\left(\frac{k_B T}{\hbar}\right)^{4\upsilon+3}\!\!\!
\int^{\beta\omega_D}_0\!\! x^{4\upsilon+2}\dfrac{e^{x}}{(e^{x}-1)^2}dx.\nonumber
\end{align}
In the last line, we define $x=\beta\:\omega$ where $\beta={\hbar}/({k_B T})$.

Now, we suppose $\hbar\,\omega\gg k_B T$, which implies that the integral in the above equation does not explicitly depend on temperature for low T. Thus, in the limit of $ \beta\,\omega_{D}\rightarrow \infty $, we have
\begin{align}
\int^{\infty}_0 x^{4\upsilon+2}e^{-x}dx= (4\upsilon+2)!.
\end{align}
In the high-temperature limit,where $\hbar\,\omega\ll k_B T$, the ratio of exponential functions in Eq.~(44) simplifies to $x^{-2}$, thus we obtain
\begin{align}
\int^{\beta\omega_D}_0 x^{4\upsilon+2}x^{-2} dx= \dfrac{(\beta\,\omega_D)^{4\upsilon+1}}{4\upsilon+1}.
\end{align}
Thus, the master equation becomes
\begin{align}
\dot{\rho}_s(t)=-\gamma\Big((S^z)^2 \rho_s (t)+\rho_s (t)(S^z)^2 -2(S^z)^2 \rho_s (t)(S^z)^2\Big),
\end{align}
where the decay rate is
\begin{align}
\gamma = 2\pi \dfrac{8\pi^2 A^2 \lambda^2_{00} }{\nu^4_{s}(\omega_D)^{4\upsilon}}\left( \frac{k_B T}{\hbar}\right) ^{4\upsilon+3} (4\upsilon+2)! ,
\end{align}
for $\hbar\,\omega\gg k_B T$ 
and
\begin{align}
\gamma = 2\pi \dfrac{8\pi^2 A^2 \,\lambda^2_{00}\, \omega_D }{\nu^4_{s}(4\upsilon+1)}\left( \frac{k_B T}{\hbar}\right) ^{2},
\end{align}
for $\hbar\,\omega\ll k_B T$. 

%
%
\bibliography{dec}

\end{document}